\documentclass[preprint,showkeys,preprintnumbers,amsmath,amssymb]{revtex4}

\usepackage{graphicx}
\usepackage{dcolumn}
\usepackage{bm}

\begin{document}

\preprint{APS/123-QED}

\title{In-situ comprehensive calibration of a tri-port nano-electro-mechanical device}

\author{E. Collin} 
\email{eddy.collin@grenoble.cnrs.fr}
\author{M. Defoort}
\author{K. Lulla}
\author{T. Moutonet}
\author{J.-S. Heron}
\author{O. Bourgeois}
\author{Yu. M. Bunkov}
\author{H. Godfrin}

\affiliation{%
Institut N\'eel
\\
CNRS et Universit\'e Joseph Fourier, \\
BP 166, 38042 Grenoble Cedex 9, France \\
}%

\date{\today}

\begin{abstract}
We report on experiments performed in vacuum and at cryogenic temperatures on a tri-port nano-electro-mechanical (NEMS) device. One port is a very non-linear capacitive actuation, while the two others implement the magnetomotive scheme with a linear input force port and a (quasi-linear) output velocity port. 
We present an experimental method enabling a full characterization of the nanomechanical device harmonic response: the non-linear capacitance function $C(x)$ is derived, and the normal parameters $k$ and $m$ (spring constant and mass) of the mode under study are measured through a careful definition of the motion (in meters) and of the applied forces (in Newtons). 
These results are obtained with a series of purely electric measurements performed without disconnecting/reconnecting the device, and rely only on known DC properties of the  circuit, making use of a thermometric property of the oscillator itself: we use the Young modulus of the coating metal as a thermometer, and the resistivity for Joule heating. The setup requires only three connecting lines without any particular matching, 
enabling the preservation of a high impedance NEMS environment even at MHz frequencies. 
The experimental data are fit to a detailed electrical and thermal model of the NEMS device, demonstrating a complete understanding of its dynamics.
These methods are quite general and can be adapted (as a whole, or in parts) to a large variety of elecromechanical devices.
\end{abstract}

\keywords{nano-mechanics, electro-mechanics, dynamics, non-linearities, low temperatures}
\maketitle

\section{INTRODUCTION}

The field of micro and nowadays nano-mechanics is continuously expanding, presenting  applications in a variety of topical fields including biophysics and chemistry (for e.g. microfluidics \cite{fluids}, chemical sensing \cite{sensing1,sensing2,chemical,bio}, attogram mass sensing \cite{mass1,mass2}).

In many applications, the input and output signals are obtained in arbitrary units without hindering the use of the device. 
However, in some cases it is {\it essential} to know the applied forces in Newton and the displacements in meters: e.g. obviously for an absolute position or force sensor, but also for a physical characterization and understanding of these mechanical systems themselves. For low frequency micro-electro-mechanical devices (MEMS, typically 10$~$kHz), the connecting lines and the on-chip wiring behave essentially DC: the voltage applied with the generator, or detected with the amplifier is truly the voltage on the MEMS. For opto micro or nano-mechanical devices (MOMS or NOMS), the output is usually directly obtained in meters: the difficulty is (only) technical and lies in the ability of bringing the optics {\it on} the chip (by focusing the beam, and avoiding heating).\\
For nano-electro-mechanical oscillators (NEMS) resonating above 10$~$MHz and up to GHz frequencies, the electric connections are of a concern: attenuation will occur in lossy cables, line resonances may appear due to discrete/distributed $LC$ contributions, and impedance mismatches will generate complex reflection patterns. If no care is taken, the voltage at one end of the cables is {\it certainly not} the same as the one on the other side, on the NEMS. 
On the other hand, if particular care is taken a 50$~\Omega$ matched circuitry can be set up; this is the optimal design as far as the {\it voltage signal amplitude} is concerned. \\
However, high impedance environments are required when {\it isolating} the NEMS is essential. 
 This is the case when addressing the fine study of loss mechanisms at low temperatures \cite{jeevakQ1,jeevakQ2,OwersB}. Many low temperature experiments (including our devices) use the practical magnetomotive scheme, exciting one resonance of the mechanical structure \cite{clelandroukes,OwersB,qfs}. Unfortunately, a (low) impedance load $Z_l(\omega)$ in a (high) magnetic field $B$ contributes to the overall measured damping mechanism \cite{OwersB,clelandQ}, with corrections to both the resonance position and the linewidth  proportional to $Im(Z_l)\, B^2/\left| Z_l \right|^2 $ and $Re(Z_l)\, B^2/\left| Z_l \right|^2 $, respectively.\\
Isolating the device is also crucial for the fine study of quantum nanomechanics \cite{quantum1,quantum2}. With a strong coupling to the environment (the driving field), it is indeed possible to cool down the device into its ground state when using {\it feedback cooling schemes} \cite{naik,teufel}.
However, the device is then kept in a strongly {\it out-of-equilibrium} state where only one degree of freedom is cold, the one under active pumping. The scheme is clearly impractical when studying fundamental coherence properties on their own, which require a true {\it in-equilibrium} situation. \\
Finally, in some situations the practical realization of a 50$~\Omega$ match may prove to be difficult. A procedure enabling the verification of the actual  matching obtained at the NEMS level is thus important.
Moreover, when keeping deliberately a high impedance environment on chip, the line transmission characteristics are almost certainly unpredictable and need calibration; although not optimal signalwise, the calibrated high-impedance configuration is {\it perfectly functional}.

Many experimental setups use an electromotive drive scheme (see for e.g. Refs. \cite{jeevakQ2,nonlinC2}). For actuation, the device is coupled to an electrode $A$ through a position-dependent capacitance $C_A(x)$. The detection is obtained through another electrode $D$ with coupling capacitance $C_D(x)$. By definition, this scheme is highly nonlinear. In practice with complex electrode patterns, the capacitance functions $C_i(x)$ are very difficult to obtain. One can make estimations with crude analytic formulas, or use sophisticated finite element simulations (see e.g.  \cite{nonlinC1,nonlinC2,nonlinC3}). But {\it in fine}, only the measurements can tell what capacitance has been effectively implemented on chip.

In the present paper we report on experiments performed on a tri-port NEMS oscillator resonating around 7$~$MHz. 
One port is a nonlinear capacitive drive, another port is a linear magnetomotive drive while the last port is an almost linear detection port. Since the magnetomotive drive requires rather high magnetic fields, the experiments are performed at 4.2$~$Kelvin, in vacuum (less than 10$^{-6}~$mbar) and under about 1 Tesla.\\
We present methods enabling the characterization of the two input force ports, and of the output port. The capacitance of the nonlinear port is described in terms of a Taylor series which coefficients are obtained experimentally. The measurements are purely electric and performed without disconnecting/reconnecting the device. They rely only on DC properties, and allow {\it any} type of wiring to be used: in particular, no impedance matching is required and high-impedance termination is supported. 
The actual transmission characteristics of the lines are deduced as a function of the frequency.
The experimental results are fit to a {\it detailed analytic modeling} of the moving device, demonstrating a complete understanding of its dynamics.

\section{SETUP}
\label{explain}

\begin{figure}[t!]
\includegraphics[height=12 cm]{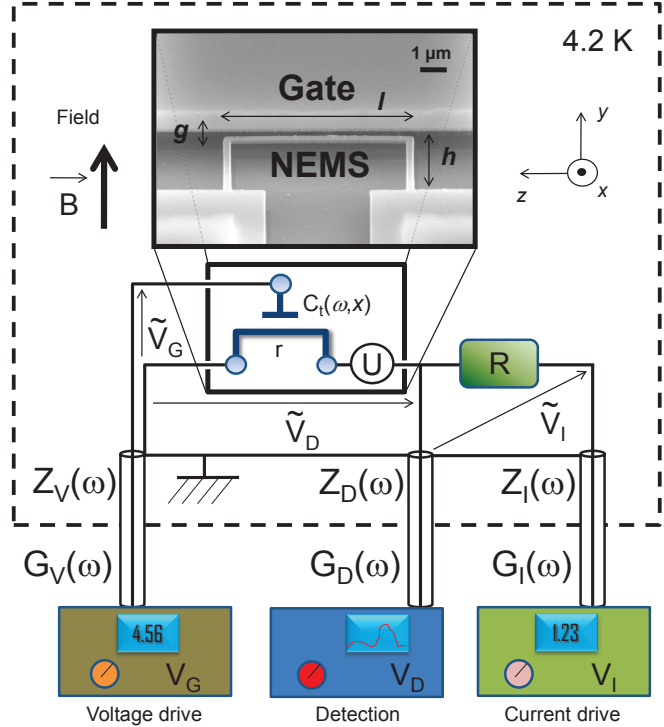}
\caption{\label{setup} (Color online) 
Schematic drawing of the experimental setup, with Scanning Electron Microscope (SEM) image of the sample (unit reference trihedron also shown). The cold part is the dashed box (chip and bias resistor $R$ only). The NEMS itself has an electric resistance $r$ and the gate electrode creates a coupling capacitance $C_t(\omega,x)$. The source $U$ represents the induced voltage under motion. Black connecting lines represent short twisted copper wires. Three coaxial cables (cylinders) link the tri-port device to conventional room temperature electronics. The lines transmission coefficients are named $G_I(\omega)$, $G_V(\omega)$ and $G_D(\omega)$ for the magnetomotive drive port, the gate capacitance port and the detection port respectively (tilded are on-chip voltages). The input impedances of the corresponding room temperature instruments as seen from the NEMS are named $Z_\lambda(\omega)$ with $\lambda=I,V,D$. }
\end{figure}

The experimental setup is drawn in Fig. \ref{setup}, and has been presented in Ref. \cite{qfs}. 
It consists of a goal-post shaped NEMS device (two "feet" of length $h=3.1~\mu$m plus a "paddle" of $l=7~\mu$m, 280$~$nm wide) fabricated from a thick SOI substrate. The oxide layer was 1$~\mu$m thick for a top silicon thickness thinned down to 150$~$nm (low-doped P type, resistivity at room temperature of 14-22$~\Omega . $cm). The structure is patterned by means of Reactive Ion Etching (SF$_6$ plus O$_2$ plasma) through an Al sacrificial mask obtained by e-beam lithography. The beams are released after HF chemical vapor etching. About 30$~$nm of aluminum has been deposited on top of the structure to create electrical contacts. We used thermal evaporation, with the chip kept at room temperature in a vacuum of about $10^{-6}~$mbar. This metallic layer is rather soft, leaving the structure unstressed (no measurable room-temperature bending) and adding only little elasticity \cite{coatings}. The low-teperature distortion appearing from differential thermal expansion (bimorph effect) is estimated to be at most of the order of the thickness of the structure. 

The mobile part is composed of the two cantilever feet linked by the paddle (Fig. \ref{setup}). Due to symmetry, in its first resonant mode (out-of-plane flexural mode) the device behaves as a simple cantilever loaded by half the paddle at its extremity. This geometry has been studied extensively in our group, from MEMS to NEMS scales with various metallic coatings \cite{qfs,JLTP_VIW,qfs2009,coatings,nonlinPRB}. In the following, the quoted displacement $x$ corresponds to the motion of the paddle (top end of the cantilevers).
The sample is placed within a helium cryostat, in a small cell in which cryogenic vacuum is maintained. 
The DC electric resistance $r$ of the NEMS is 110$~\Omega$ at 4.2$~$K using a 4-wire measurement \cite{qfs}.

The device has two actuation ports, and one detection port that we use to study the harmonic response of the first mode at frequency $\omega_0$ (Fig. \ref{setup}).
One drive port creates a linear force (i.e. independent of the displacement $x$) by means of the magnetomotive scheme: a current $I(t)=I_0 \cos (\omega t)$ is fed through the metallic layer covering the suspended part with a cold bias resistor (voltage $\tilde{V}_I(t)$ across $R+r=1.1~$k$\Omega$). In a static magnetic field $\vec{B}$ oriented along the feet of the moving structure, the resulting force is perpendicular to the chip surface and writes $F(t)=I_0 l B \cos (\omega t)$. The studied harmonic response simply writes $x(t)=x_0 \cos(\omega t+\varphi)$.\\
The detection of the motion is carried out through the measurement of the induced voltage $U(t) = \tilde{V}_D(t) - r I(t)$ appearing at the structure's ends while it moves and cuts the field lines. This output voltage can be shown to be {\it almost linear}, reducing to the expression $U(t) = l B \dot{x}(t)$ \cite{JLTP_VIW,note}. By sweeping $\omega$ around $\omega_0$, a resonance peak in $\tilde{V}_D$ is detected.\\
The second drive port is realized with a gate electrode capacitively coupled to the moving structure. The gap between the NEMS and the gate is about $g \approx 100~$nm.
The total gate capacitance $C_t$ can be written:
\begin{displaymath}
C_t = C_0 +  \int_{NEMS} \delta C,
\end{displaymath}
with $C_0$ the capacitance due to the leads and connecting pads, and $\int_{NEMS} \delta C$ is the NEMS metallic layer contribution. Since the capacitance falls to zero extremely quickly with the distance to the gate electrode, in the latter integral we can keep only the paddle contribution.  
Parameterizing the NEMS capacitance with the global position of the paddle in the $\vec{x}$ and $\vec{y}$ directions (see Fig. \ref{setup}) leads to:
\begin{displaymath}
 \int_{NEMS} \delta C =C(x,y).
\end{displaymath}
The potential energy originating from the voltage bias $\tilde{V}_G (t)$ on the gate electrode writes then:
\begin{displaymath}
E_C = \frac{1}{2} C(x,y) \tilde{V}_G^2.
\end{displaymath}
The resulting force (on the mobile part) is thus:
\begin{displaymath}
\vec{F}_C = +\frac{1}{2} \left( \frac{\partial C(x,y)}{\partial x} \vec{x} + \frac{\partial C(x,y)}{\partial y} \vec{y} \right) \tilde{V}_G^2.
\end{displaymath}
Static deflections that are generated by a DC gate voltage (or a DC magnetomotive current) can be safely neglected; only the harmonic response has to be considered, and the modeling presented does not incorporate any static distortions. 
In experiments, the displacements $x,y$ remain small. We thus proceed with the following Taylor series expansions:
\begin{eqnarray*}
\frac{\partial C(x,y)}{\partial x} & = & \frac{\partial C(0,0)}{\partial x}+\frac{\partial^2 C(0,0)}{\partial x^2} x \\
&+ & \frac{1}{2} \frac{\partial^3 C(0,0)}{\partial x^3} x^2 + \frac{\partial^2 C(0,0)}{\partial x \partial y} y \\
&+ & \frac{\partial^3 C(0,0)}{\partial x^2 \partial y} x y + \frac{1}{6} \frac{\partial^4 C(0,0)}{\partial x^4} x^3 ,  \\
\frac{\partial C(x,y)}{\partial y} & = & \frac{\partial C(0,0)}{\partial y}+\frac{\partial^2 C(0,0)}{\partial x \partial y} x \\
&+ & \frac{1}{2} \frac{\partial^3 C(0,0)}{\partial x^2 \partial y} x^2 + \frac{\partial^2 C(0,0)}{\partial y^2} y \\
&+ & \frac{\partial^3 C(0,0)}{\partial x \partial y^2} x y  + \frac{1}{6} \frac{\partial^4 C(0,0)}{\partial x^3 \partial y} x^3,
\end{eqnarray*}
where we kept only orders smaller than 3. The $y$ displacement is at lowest order an $x^2$ term \cite{JLTP_VIW,note}; we estimate $y \approx -  \frac{3}{5} \, x^2/h$.
Obviously, this description breaks down for $x>\!>g$: higher orders should be taken into account making the capacitance $C(x,y)$ fall smoothly to zero.

The $\vec{y}$ component of $\vec{F}_C$ (denoted $F^y_C$) is an axial force load acting on the feet of the structure. It influences their spring constants $k_{foot}$, and we have at first order \cite{JLTP_VIW,coatings}:  
\begin{displaymath}
k_{foot} = \frac{k_0}{2} \left( 1 - \phi \frac{F^y_C/2 \, h^2}{E \, I_y} \right) 
\end{displaymath}
with $k_0/2$ the unaxially-loaded spring constant of each foot, $E$ the corresponding Young modulus and $I_y$ its second moment of area. $\omega_0=\sqrt{k_0/m_0}$ by definition, with $m_0$ the normal mass of the mode.
$\phi$ is a small (mode dependent) number that is estimated for our geometry to be about $+0.095$. The end mass load due to the paddle is taken into account in this writing \cite{coatings}. 

The $\vec{x}$ component $F^x_C$ of the force generates directly a modulation of the NEMS restoring force. Combining the axial load $F^y_C$ with the latter, we finally obtain the {\it effective} out-of-plane gate contribution:
\begin{eqnarray}
F_C^{ef\!f} & = & +\frac{1}{2} \frac{\partial C(0,0)}{\partial x}  \tilde{V}_G^2 \nonumber \\
&  &\!\!\!\!\!\!\!\!\!\!\!\!\!\!\!\!\!\!\!\!\!\!  + \frac{1}{2} \left[ \frac{\partial^2 C(0,0)}{\partial x^2}  +\left(  \frac{\phi}{2} \frac{k_0  h^2}{E I_y} \right) \frac{\partial C(0,0)}{\partial y} \right]\, x\,  \tilde{V}_G^2 \nonumber \\
&  & \!\!\!\!\!\!\!\!\!\!\!\!\!\!\!\!\!\!\!\!\!\!\!\!\!\!\!\!\!  + \frac{1}{2} \left[ \frac{1}{2} \frac{\partial^3 C(0,0)}{\partial x^3}+ \left( \frac{\phi}{2} \frac{k_0  h^2}{E I_y} - \frac{3}{5}\frac{1}{h}\right) \frac{\partial^2 C(0,0)}{\partial x \partial y}  \right]\, x^2\,  \tilde{V}_G^2 \nonumber \\
&  & \!\!\!\!\!\!\!\!\!\!\!\!\!\!\!\!\!\!\!\!\!\!\!\!\!\! + \frac{1}{2} \left[ \frac{1}{6} \frac{\partial^4 C(0,0)}{\partial x^4} + \left( \frac{\phi}{4} \frac{k_0  h^2}{E I_y} - \frac{3}{5}\frac{1}{h}\right) \frac{\partial^3 C(0,0)}{\partial x^2 \partial y} \right. \nonumber \\
& & \left.  \!\!\!\!\!\!\!\!\!  -\left( \frac{\phi}{2} \frac{k_0  h^2}{E I_y} \frac{3}{5}\frac{1}{h} \right)  \frac{\partial^2 C(0,0)}{\partial y^2} \right]\, x^3\,  \tilde{V}_G^2 \label{capa}.
\end{eqnarray}
The spatial variation of the capacitance $C(x,y)$ occurs on a typical lengthscale of the order of the electrode's gap $g$. Therefore, the terms in Eq. (\ref{capa}) between parenthesis should be compared to this parameter. We obtain $ \frac{\phi}{2} \frac{k_0  h^2}{E I_y}g \approx 0.01$ and $\frac{3}{5}\frac{1}{h} g \approx 0.02$, and conclude that for our device the corresponding factors in Eq. (\ref{capa}) can be safely neglected: only the $x$ variation of the NEMS capacitance $C$ has to be considered. Note however that the formalism given below remains perfectly valid in the more general case where the tension contribution $F^y_C$ is not negligible, using the full coefficients of Eq. (\ref{capa}). For a given sample, it is rather difficult to predict accurately the actual values of the $\partial^n C(0,0)/\partial x^{n-p}\partial y^p$ ($n,p=1,2,3,4$).

In practice, the experimentalist applies biases $V_I$, $V_G$ and detects a voltage $V_D$ (Fig. \ref{setup}). These signals are related to the true on-chip 
values $\tilde{V}_I$, $\tilde{V}_G$ and $\tilde{V}_D$ through the line transmission coefficients defined as:
\begin{eqnarray*}
G_I (\omega) & = & \frac{\tilde{V}_I}{V_I}, \\
G_V (\omega) & = & \frac{\tilde{V}_G}{V_G}, \\
G_D (\omega) & = & \frac{\tilde{V}_D}{V_D}.
\end{eqnarray*}
The frequency-dependence of these coefficients arises from experimental imperfections (distributed $RLC$ lines, impedance mismatches).
Similarly, the (equivalent) coupling capacitance $C_0$ writes $C_0(\omega)$, because in practice the connecting pads capacitance is combined to the inductance of the bonding wires (and more generally, to the finite impedance of the close-by wiring). Note however that since its size is vanishingly smaller than the wavelength of the electric signals under use, the NEMS contribution to $C_t$ should be frequency-independent. We thus finally write:
\begin{eqnarray}
C_t(\omega,x) & = & C_0 (\omega) + C(x). \nonumber
\end{eqnarray}
Note that in experimental implementations $C(x) <\!< C_0$. The capacitance $C_0$ is responsible for a current $C_0 \, d \tilde{V}_G/dt$ flowing through the NEMS, adding up to the current used for the magnetomotive drive.

The low frequency (or DC) values are always well-defined, and by definition $G_I (0)=G_V (0)=G_D (0)=1$, together with $C_0(0)=C_{00}$. 
By measuring the current $V_D/r$ in the NEMS under a small low frequency voltage $V_G$ on the gate, we obtain $C_{00} = 0.32~$pF \cite{qfs}.
However, depending on the impedance matching of the connecting lines and on the quality of the wiring, the frequency dependence of the above coefficients can be very complex and unpredictable in the MHz - GHz range. 

In the following we present two distinct techniques enabling the full characterization of the electromechanical device.
In a first step, the calibration procedure described in Sec. \ref{experiments} defines experimentally the spatial-dependence of the capacitance $C(x)$ (up to about $x \approx g$). It is obtained through the measurement of the series coefficients $\partial^n C(0,0)/\partial x^n$ ($n=1,2,3,4$) appearing in Eq. (\ref{capa}). \\
In a second step, another experimental calibration procedure gives access to the frequency-dependences $G_\lambda(\omega)$ ($\lambda=I,V,D$) and $C_0(\omega)$ (here, up to 30$~$MHz). It relies only on a thermal property which is essentially DC. 
A comprehensive thermal model is developed and compared to the experimental data, proving excellent agreement without free parameters.

\section{EXPERIMENT}
\label{experiments}

In practice, the two generators in Fig. \ref{setup} ($V_G$ and $V_I$) are implemented by a Tektronix AFG3252 dual channel arbitrary waveform generator.
The detection of $V_D$ is realized with a Stanford SR844 RF lock-in amplifier, giving access to both the in-phase $X$ and out-of-phase $Y$ components  of the harmonic motion (homodyne detection). The magnetic field is obtained with a small superconducting coil and a Kepco 10$~$A DC current source. All presented data are obtained at 4.2$~$K in a field of 840$~$mT, but the thermal variation of the mechanical parameters has been carefully studied from 4.2$~$K to 30$~$K. The expected magnetic field dependence has been verified from about 145$~$mT up to 1.1$~$T. On-chip applied voltages never exceed 10$~$V peak, and currents through the NEMS are kept below $200~\mu$A peak. 
Note that experimental results are quoted in {\it root-mean-square} values.

\subsection{Spatial-dependence of $C(x)$}
\label{spatial}

From Eq. (\ref{capa}), we immediately realize that the applied voltage $\tilde{V}_G$ allows to:
\begin{itemize}
\item drive the NEMS' mode under study, through the first term $\partial C(0,0)/\partial x$ and a voltage $\tilde{V}_{G0} \cos(\omega' t)$ oscillating around $\omega_0/2$,
\item tune the resonance frequency, with the second term $\partial^2 C(0,0)/\partial x^2$ and a DC voltage bias $V_G$,
\item and adjust the Duffing-like nonlinearity, with the last terms $\partial^3 C(0,0)/\partial x^3$, $\partial^4 C(0,0)/\partial x^4$ and a DC bias.
\end{itemize} 
Inversely, measuring these effects we can quantify the parameters $\partial^n C(0,0)/\partial x^n$ ($n=1,2,3,4$).
These NEMS capacitive frequency tuning and nonlinearity tuning have been experimentally demonstrated for the first time in Ref. \cite{tuningRoukes} with a doubly clamped beam.
The formalism is more complex since the capacitance terms have to be integrated over the  mode distortion \cite{tuningRoukes}, introducing additional shape factors that have to be calculated; however, the final conclusions are strictly identical.
Note that using the resonance frequency tuning, parametric drive \cite{rugarPRL} can be implemented with a proper choice of gate voltage modulation \cite{qfs,PRBtobe}.

\begin{figure}[t!]
\includegraphics[height=9 cm]{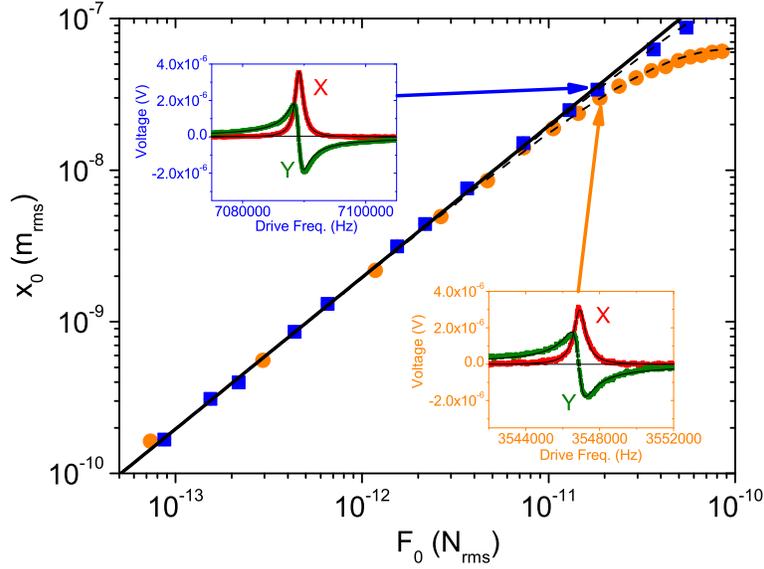}
\caption{\label{drives} (Color online) Displacement versus force curves for the magnetomotive drive (blue squares) and the capacitive drive (orange circles).
Dashed lines are fits showing a saturation at large $x$ (with respect to the straight full line). Values are quoted in {\it rms}, and error bars are typically $\pm 5~$\%, the size of symbols. Inset: Measured resonance lines (X and Y components) with Lorentzian fits (black lines). Note the wide dynamic range explored. }
\end{figure}

In Fig. \ref{drives} we present the characteristic displacement versus force curves obtained for the magnetomotive drive and the capacitive drive. The axis units have been calculated from the magnetomotive drive using the careful calibrations of $G_I(\omega)$ and $G_D(\omega)$ presented in Sec. \ref{frequency}.
Knowing $G_V(\omega)$, scaling the electromotive force $+\frac{1}{2} \frac{\partial C(0,0)}{\partial x}  \frac{\tilde{V}_{G0}^2}{2} \cos(2\omega' t)$ on the magnetomotive force $I_0 l B \cos(\omega t)$, for the same displacement obtained in the linear regime, gives access to $\partial C(0,0)/\partial x$; we obtain $+2.1 \times 10^{-11}~$F/m.

For small deflections (typically $x<30~$nm$_{rms}$), both excitation methods give a Lorentzian response in the detected induced voltage (insets, Fig.  \ref{drives}). 
The displacement amplitude $x_0$ verifies then the simple relation: $x_0=F_0\, Q/k_0$ with $Q=f_0/\Delta f$ the quality factor of the resonance, and $\Delta f$ its full width at half height (FWHH) measured on $X$ (in-phase). We obtain $Q=5.1 \times 10^3$ (i.e. $f_0=7.09~$MHz and $\Delta f=1.4~$kHz), $k_0=2.55~$N/m and $m_0=1.3\,10^{-15}~$kg which fully characterizes the oscillator in the linear regime. Spring constant and mass are in excellent agreement with calculated parameters taking into account the end mass load and the coating metal (typically $\pm 5~$\%, formalism from Refs. \cite{JLTP_VIW,coatings}).\\ 
However, for large deflections the device becomes nonlinear (see dashed lines in Fig. \ref{drives}). With the magnetomotive drive scheme, the resonance line remains Lorentzian-looking up to very large $x_0$, demonstrating very small Duffing-like contributions in the dynamics equation \cite{qfs}: writing the restoring force $F_{restore}=-k_0(1+k_1 x+k_2 x^2)\,x$, we have $k_1\,x,k_2\,x^2 <\!<1$. But the linewidth grows with the displacement while the resonance shifts down, a signature of both {\it anelasticity} and {\it Joule heating} \cite{qfs,coatings}. \\
With the electromotive drive scheme, these intrinsic nonlinear behaviors are superimposed to voltage-dependent nonlinearities. Since $[\tilde{V}_{G0} \cos(\omega' t)]^2=\tilde{V}_{G0}^2[1+\cos(2\omega' t)]/2$, a static term $\tilde{V}_{G0}^2/2$ appears in Eq. (\ref{capa}) which is responsible for a negligible static deflecting force (through $\partial C(0,0)/\partial x$), a small frequency shift (through $\partial^2 C(0,0)/\partial x^2$), and finally a small nonlinear Duffing-type contribution ($\partial^3 C(0,0)/\partial x^3$,$\partial^4 C(0,0)/\partial x^4$). Moreover, the $\cos(2\omega' t)$ drive component in the series is also nonlinear with the same coefficients. The resulting resonance line at large drives looks like a "pointed hat" on $X$, with a height saturating around $65~$nm$_{rms}$. This saturation is characteristic of the decrease towards zero of $C(x)$ at large deflections.
\begin{figure}[t!]
\includegraphics[height=9 cm]{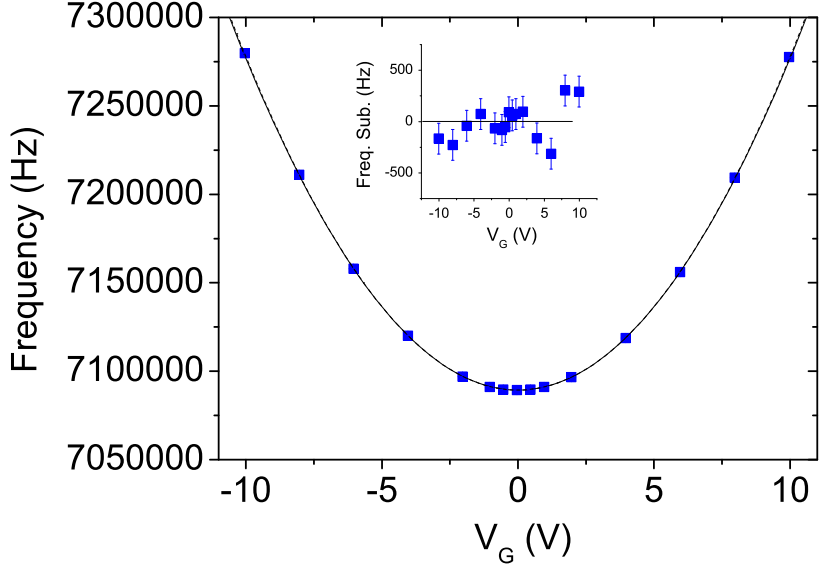}
\caption{\label{freq} (Color online) Frequency tuning of the resonance $\omega_{res}/2 \pi$ using a DC voltage on the gate. For this measurement, the device motion has been kept in the linear regime ($x<7~$nm$_{rms}$). 
The dashed line is a quadratic function while the full line also takes into account a small $V_G^4$ component, recalculated from the fit of Fig. \ref{nonlindep}. Both lines are practically indistinguishable one from the other.
Inset: after subtraction of the full expression $+1850 \, V_{G}^2 -0.1 \, V_{G}^4$ (a small DC offset voltage has also been taken into account). Error bars about $\pm150~$Hz. Note the deviations arising at large biases, due we believe to the activation of offset charges in the sample. }
\end{figure}

\begin{figure}[t!]
\includegraphics[height=9 cm]{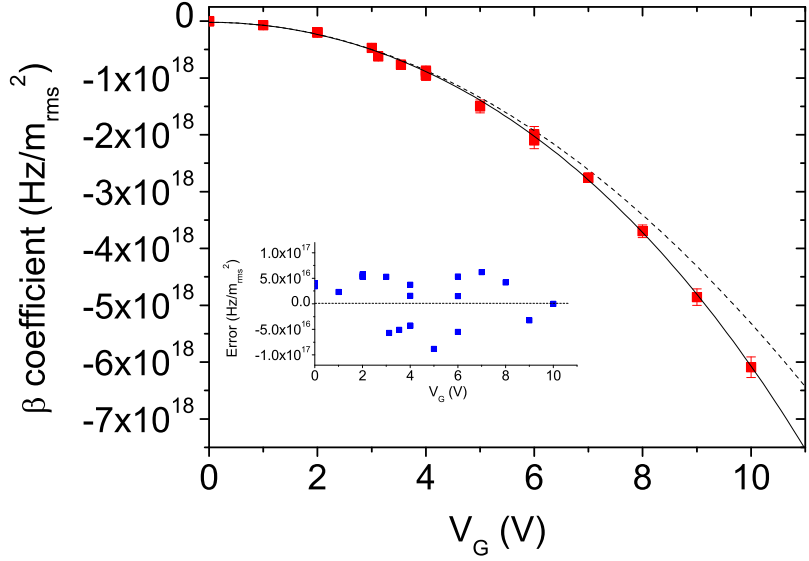}
\caption{\label{nonlindep} (Color online) Nonlinear frequency pulling coefficient $\beta$ as a function of the DC gate voltage. 
The dashed line is a quadratic function, while the full line is the complete fit, yielding $-5.5 \times 10^{16} \, V_{G}^2-7.5 \times 10^{13} \, V_{G}^4~$Hz/m$_{rms}^2$.
}
\end{figure}

Keeping the NEMS in the linear regime with the magnetomotive drive scheme ($x<7~$nm$_{rms}$), we measure the frequency tuning using a DC voltage on the gate $V_{G}$ (Fig. \ref{freq}). 
Most of the effect arises from the gate modulation of the spring constant ($\partial^2 C(0,0)/\partial x^2$). However to be perfectly rigorous, the gate bias creates a static deflecting force which can be taken into account as well. If the intrinsic restoring force of the device is slightly nonlinear with $F_{restore}=-k_0(1+k_1 x+k_2 x^2)\,x$, the static deflection combined with the $k_1$ term contributes at first order to the frequency shift. We neglect its second order contribution (which enters the small $V_{G}^4$ term below).
Solving the dynamics equation, we obtain for the Lorentzian response a resonance position:
\begin{eqnarray}
\omega_{res} &=& \omega_0 \nonumber \\ 
& + & \omega_0 \left( \frac{ k_1 \frac{\partial C(0,0)}{\partial x}- \frac{1}{2}  \frac{\partial^2 C(0,0)}{\partial x^2} }{2 k_0} \right) V_{G}^2 \nonumber \\ 
& + & \omega_0 \left( \frac{-\left(  k_1 \frac{\partial C(0,0)}{\partial x} - \frac{1}{2}  \frac{\partial^2 C(0,0)}{\partial x^2} \right)^2+2  \frac{\partial C(0,0)}{\partial x} \left(   k_1 \frac{\partial^2 C(0,0)}{\partial x^2} - \frac{1}{2} \frac{\partial^3 C(0,0)}{\partial x^3} \right)}{8 k_0^2} \right) V_{G}^4  ,  \label{freqshift}
\end{eqnarray}
where we developed the gate voltage dependence to order $V_{G}^4$. 
Knowing $\omega_0$ and $k_0$ from the preceding paragraph, and neglecting $k_1$ (the device is naturally very linear), we obtain at first order the quadratic dependence of Fig. \ref{freq}. The fit brings $\partial^2 C(0,0)/\partial x^2=-0.0027~$F/m$^2$.
In the inset, 
Fig. \ref{freq} we plot the deviation to the total fit, taking into account a small $V_G^4$ term recalculated from parameters obtained in the following section (Eq. (\ref{beta}) and fit in Fig. \ref{nonlindep}). 
A small DC offset voltage has also been taken into account (smaller than 100$~$mV amplitude). This contribution seems to be on chip, arising from local stray electric fields. At the same time, above about 8$~$V (absolute), deviations appear which display also an hysteretic behavior. Both DC gate offset and large voltage deviations are cool-down dependent.
We believe that they are due to offset charges in the aluminum-coated silicon sample, which can be activated above a given voltage/temperature threshold \cite{mike}.  
\begin{figure*}
\includegraphics[height=7cm]{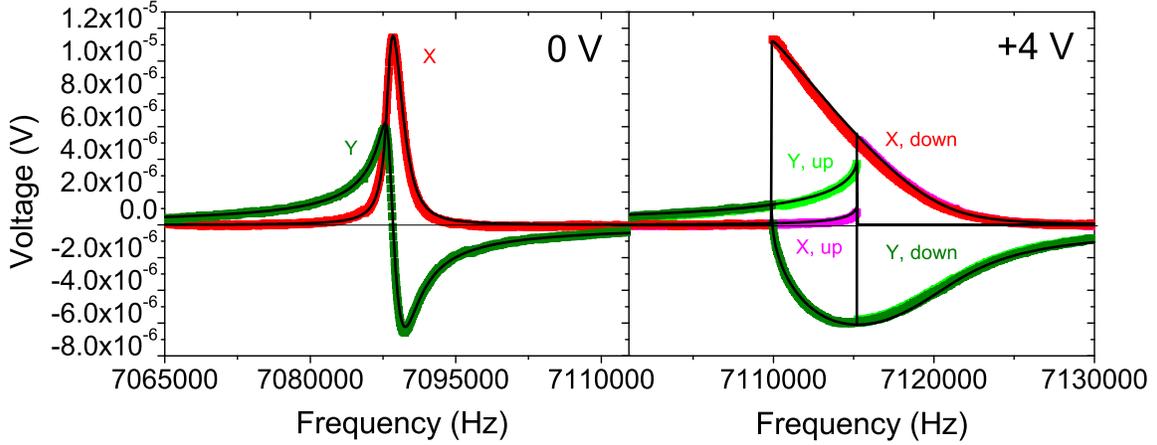}
\caption{\label{nonlinlines} (Color Online) Resonance lines obtained for an excitation of 73$~$pN$_{rms}$ (max. amplitude of motion 110$~$nm$_{rms}$). Left graph, without any bias on the gate electrode. Right graph, with +4$~$V DC on the gate, the line becomes strongly nonlinear and displays hysteresis \cite{landaumeca}. On both graphs, the black lines are fits based on Refs. \cite{coatings, nonlinPRB}. }
\end{figure*}

For large DC gate biases and large deflections, the device becomes nonlinear (Fig. \ref{nonlinlines}).
The gate voltage-dependent terms of Eq. (\ref{capa}), by adding up to the force $F_{restore}=-k_0(1+k_1 x+k_2 x^2)\,x$, generate a frequency pulling factor $\beta$:
\begin{equation}
\omega_{pos} = \omega_{res} + \beta(V_{G}) \, x_0^2 \label{freqequa}.
\end{equation}
The resonance line becomes bistable above a given displacement threshold, and $\omega_{pos}$ represents the measured position of the $X$ peak maximum when sweeping the magnetomotive drive frequency $\omega$ in the direction of the pulling (downwards for $\beta<0$ and upwards for $\beta>0$) \cite{landaumeca}.
Using the nonlinear coefficients of Ref. \cite{nonlinPRB}, we obtain in the high-$Q$ limit:
\begin{eqnarray}
\beta(V_{G}) & = & \beta_0 \nonumber \\
& + & \omega_0  \left[ \frac{\left( \frac{1}{9} k_1^3 - 3 k_1\,k_2\right) \frac{\partial C(0,0)}{\partial x} - \frac{1}{18} k_1^2 \frac{\partial^2 C(0,0)}{\partial x^2} + \frac{5}{6} k_1 \frac{\partial^3 C(0,0)}{\partial x^3}- \frac{1}{8} \frac{\partial^4 C(0,0)}{\partial x^4} }{4 k_0} \right] V_{G}^2 \nonumber \\ 
& + & \omega_0  \left[ - \frac{ \frac{5}{24} \left( \frac{\partial^3 C(0,0)}{\partial x^3} \right)^2}{8 k_0^2} + \frac{\frac{1}{54} k_1 \frac{\partial^2 C(0,0)}{\partial x^2} }{8 k_0 } \left( \frac{ k_1 \frac{\partial^2 C(0,0)}{\partial x^2} }{k_0} + \frac{3 \frac{\partial^3 C(0,0)}{\partial x^3}}{k_0} \right)+ \right. \nonumber \\
& + &  \left. \frac{ \frac{\partial C(0,0)}{\partial x}}{8 k_0} \left( \frac{ \frac{\partial^2 C(0,0)}{\partial x^2}}{27 k_0} \left[k_1^3 - 162 \, k_1 k_2 \right] -\frac{ \frac{\partial^3 C(0,0)}{\partial x^3}}{6 k_0} \left[ k_1^2 -9 \, k_2 \right] +   \frac{k_1 \frac{\partial^4 C(0,0)}{\partial x^4}}{2 k_0}  \right) \right] V_{G}^4 \label{beta},
\end{eqnarray}
developed to order $V_{G}^4$ ($\beta_0$ being the intrinsic nonlinear coefficient, almost zero for our device). The static deflection (due to $\partial C(0,0)/\partial x$) is taken into account at first order.
\begin{figure}[t!]
\includegraphics[height=9 cm]{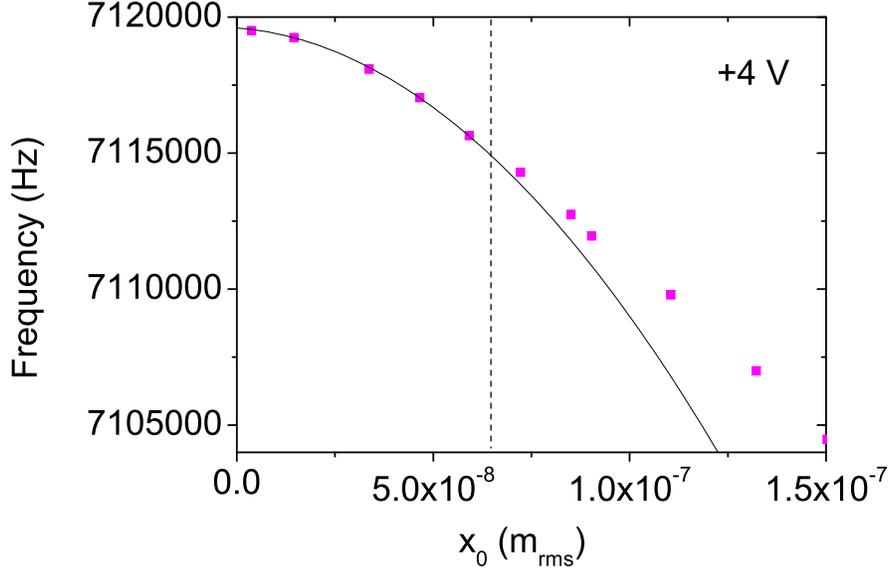}
\caption{\label{nonlinfit} (Color online) Resonance position $\omega_{pos}/2 \pi$ while sweeping the drive frequency down, as a function of the displacement. The gate voltage is biased at +4$~$V DC. The black line is a parabolic fit, yielding the value of $\beta$ (here, $-9.5 \times 10^{17}~$Hz/m$_{rms}^2$). The vertical dashed line corresponds to the saturation value of the capacitive drive, Fig. \ref{drives}, which represents a limit for the fitting procedure. Error bars size of the order of the symbols ($\pm 150~$Hz).}
\end{figure}
In Fig. \ref{nonlinfit} we present the position of the resonance measured while sweeping the frequency down, as a function of the displacement, 
for a +4$~$V DC gate bias.
At small displacements ($x<65~$nm$_{rms}$, dashed vertical), we obtain the expected quadratic dependence, Eq. (\ref{freqequa}). For larger displacements, the capacitance function $C(x)$ tends to zero and we recover the almost-linear intrinsic frequency dependence of the device (due to anelasticity and heating effects) \cite{coatings,PRBtobe}.
In Fig. \ref{nonlindep} we present the measured voltage dependence of the nonlinear coefficient $\beta$.
Most of the effect arises from $\partial^4 C(0,0)/\partial x^4$, Eq. (\ref{beta}) and leads to a quadratic dependence $V_{G}^2$. Neglecting the intrinsic nonlinear terms $k_1,k_2$ as we did in the previous paragraph, we obtain $\partial^4 C(0,0)/\partial x^4=+3.3 \times 10^{11}~$F/m$^4$.

The last capacitance coefficient which deserves to be found is $\partial^3 C(0,0)/\partial x^3$. However, its influence on the experimental parameters is rather weak: for a linear mechanical device ($k_1,k_2 \approx 0$) it appears only in the higher order $V_{G}^4$ voltage dependence, Eqs. (\ref{freqshift}) and (\ref{beta}).
For our devices (and $\left|V_G \right|< 10~$V), this high-order dependence is very small, but it can be rather large in some other nano-resonators \cite{hari}.
We obtain from the nonlinear coefficient $\beta$ (Fig. \ref{nonlindep}) the value $\partial^3 C(0,0)/\partial x^3 \approx -35\,000~$F/m$^3$, which correspond to a frequency tuning term of $-0.1\, V_G^4~$Hz, consistent with the fit of Fig. \ref{freq}.

A crude estimate of the capacitance $C(x,y)$ can be produced using a modified version of the "two parallel infinite wires" formula:
\begin{equation}
C(x,y) = \epsilon_0 \frac{\pi \, l}{\ln \left( \frac{\sqrt{(g+2 r-y)^2+(x_s-x)^2}}{r} \right) } \label{capawires},
\end{equation}
with $x_s$ an effective offset mimicking the actual dielectric assymetry of the sample, and $r$ an effective radius. This assymetry between the permittivity of vacuum and of silicon/silicon oxide is responsible for the nonzero values of $\partial C(0,0)/\partial x$,$\partial^3 C(0,0)/\partial x^3$.  
A comparison with the experimentally obtained coefficients is given in Tab. \ref{capaval}. The agreement is surprisingly good, with all parameters in accordance within about a factor of 5 at worst. The calculated value of $C(0,0)$ is 1/1000th of $C_{00}$ which validates the initial hypothesis $C(x) <\!< C_0$.
The irrelevance of the $y$-variation of $C(x,y)$ in Eq. (\ref{capa}) is corroborated as well. 
\begin{table}[h!]
\begin{center}
\begin{tabular}{|c|c|c|c|c|c|}    \hline
               & C(0,0)            & $\partial C(0,0)/\partial x$ & $\partial^2 C(0,0)/\partial x^2$ & $\partial^3 C(0,0)/\partial x^3$  & $\partial^4 C(0,0)/\partial x^4$   \\   \hline \hline
Measurement    & $\equiv$                & $+2.1 \times 10^{-11}~$F/m         & $-0.0027~$F/m$^2$                & $-35\,000~$F/m$^3$        & $+3.3 \times 10^{11}~$F/m$^4$      \\    \hline
Crude estimate & $1.8 \times 10^{-16}~$F & $+5.3 \times 10^{-11}~$F/m         & $-0.0017~$F/m$^2$                &      $-6\,500~$F/m$^3$         & $+2.0 \times 10^{11}~$F/m$^4$      \\    \hline
\end{tabular}
\caption{\label{capaval} Series coefficients of the capacitance $C(x)$. $C(0,0)$ cannot be accessed directly. For the crude modeling, chosen parameters are: $r=100~$nm, $x_s=30~$nm.}
\end{center}
\end{table}

\subsection{Frequency-dependences of $G_\lambda(\omega), C_0(\omega)$}
\label{frequency}

\begin{figure}[t!]
\includegraphics[height=9 cm]{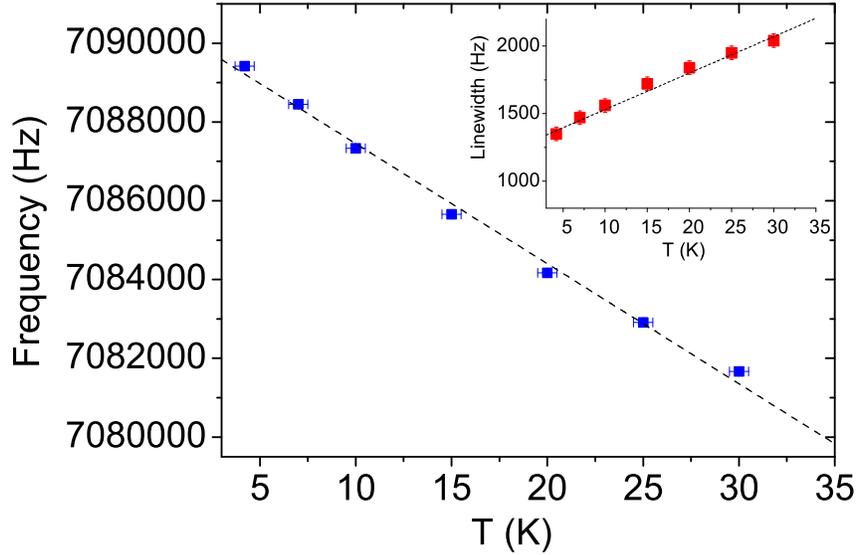}
\caption{\label{temps} (Color online) Frequency of the resonance versus temperature (the device is kept in the linear regime). Inset: corresponding linewidth as a function of $T$. For these measurements, the whole cell temperature has been regulated with a resistance bridge, from $4.2~$K up to 30$~$K. Dashed lines are linear guides (slopes -305$~$Hz/K for the frequency and +27$~$Hz/K for the linewidth respectively).}
\end{figure}

The actual voltages on-chip $\tilde{V}_\lambda$ ($\lambda=I,G,D$) can be substantially different from the applied and detected signals. In order to calibrate the setup, we measure a local property: heating through a non-resonant current $I_{heat}$. We replace the generator $V_I$ in Fig. \ref{setup} by a dual-channel generator followed by an additionner (differential amplifier with gain 1, bandwidth 100$~$MHz). We write:
\begin{eqnarray*}
V_I & = & V_{I 0} \cos(\omega t) + V_{heat} \cos(\omega'' t), \\
V_G & = & 0.
\end{eqnarray*}
The suspended part heats due to the Joule effect occurring in the aluminum metallic layer (current $\tilde{V}_I/(R+r)$ through resistance $r$, Sec. \ref{magneto}). The actual temperature of the mechanical device can be tracked by measuring both the frequency and the linewidth of the resonance, Fig. \ref{temps}. The frequency $f_0$ shifts down due to the softening of the aluminum Young modulus $E(T)$, while the dissipation (i.e. linewidth $\Delta f$) increases \cite{coatings}. \\
Heating can also be obtained with the gate electrode and a capacitive current $C_0 d \tilde{V}_G/ dt$ (Sec. \ref{capalines}). In this configuration, we apply:
\begin{eqnarray*}
V_I & = & V_{I 0} \cos(\omega t) , \\
V_G & = & V_{heat} \cos(\omega'' t).
\end{eqnarray*}
The NEMS drive voltage $V_{I0}$ (resonant at $\omega \approx \omega_0$) is kept small enough to maintain the device in its linear regime, and to produce virtually no heating.
The heating frequency $\omega''$ is chosen to be non-resonant with any of the mechanical modes of the device (even parametrically excited).
Apart from the true DC case, it is also chosen to be larger than typically $10 \Delta f$ in order to avoid nanomechanical mixing effects \cite{APLtobe}.
\\
We present below the thermal model validating our results. 
Note however that the method itself {\it does not} require the knowledge of this theory: it relies only on a scaling of the heating current amplitude at frequency $\omega''$ on the DC measured curve.
The following paragraphs deal with the drive current heating and the gate electrode current heating. Finally, the last paragraph concludes with the calibration of the detection line.

\subsubsection{Thermal model}
\label{theorthermal}

For symmetry reasons, we consider one foot of the structure loaded at the end by half of the paddle (Fig. \ref{setup}). 
The heated vibrating part is a composite beam formed by 30$~$nm of aluminum on top of 150$~$nm of monocrystalline silicon.
From the measured resistance $r$ at 4.2$~$K, we deduce the resistivity of the metallic layer: $\rho_e = 7. \times 10^{-8}~\Omega.$m, a (temperature-independent) value limited by the nanocrystalline nature of the film. 

In the Kelvin temperature range, the metal thermal conductance is solely due to the conduction electrons and can be deduced from the resistivity using the Wiedemann-Franz law \cite{kittel}: $\kappa_e =\alpha\,T~$W/m.K, with in our case $\alpha=0.35$ in S.I. units. 
For the silicon substrate, the thermal conduction is due to phonons. The phonon mean-free-path is limited by the size of the nano-bar (width $w$ and thickness $t$). Following Casimir and Ziman \cite{casimir} we write an effective conductivity: $\kappa_{ph} = 10^3 \times 1.12 \sqrt{w\,t} \, T^3~$W/m.K. This expression has been quantitatively verified on our substrates \cite{olivethermal}. 

Finally, we take for the specific heats of both materials their bulk values. For aluminum, it is mostly due to electrons and we use : $C_{V,e}=\gamma\,T~$J/kg.K with $\gamma=0.1$ in S.I. units. The silicon specific heat on the other hand is due to phonons and we have $C_{V,ph}=2.6 \times 10^{-4}\,T^3~$J/kg.K, an expression that fits measurements performed on our substrates \cite{olivecapa}.

From the numbers quoted above, we realize that the thermal properties of the metal dominate the ones of silicon from 4.2$~$K up to typically 15$~$K. Above this temperature, the phonon contribution of the silicon ($T^3$ law) gradually takes over in both the specific heat and the thermal conduction.
Heat then continuously flows from the aluminum electronic bath to the silicon phonon bath, and one needs to incorporate in the modeling the electron-phonon coupling (see e.g. Ref. \cite{jukka} for numerical values), and the Kapitza resistance at the interface \cite{kapitza}. When the structure is not heated above 15$~$K, both mechanisms remain irrelevant.

In the 4.2$~$K $-$ 15$~$K temperature range, we will thus thermally model the NEMS electronic properties only. 
The Joule heating is indeed deposited directly in the metallic layer, and we assume that the temperature of the underlying silicon substrate adiabatically follows the aluminum coating.
Since the transport is diffusive, Fourier's heat equation is used.
The heat equation for $T(y,t)$ writes then:
\begin{equation}
\rho_m \, C_V[ T(y,t) ] \frac{\partial T(y,t)}{\partial t} -\frac{\partial }{\partial y} \left[\kappa[T(y,t)] \frac{\partial T(y,t)}{\partial y} \right] = \dot{q} (y,t)  ,  \label{heat}
\end{equation}
with $C_V( T )$ the specific heat (per mass, $\rho_m$ being the metal density) and $\kappa(T)$ the thermal conductivity of the metal layer. $\dot{q} (y,t)$ is the heat load (per unit volume) at abscissa $y$ and time $t$:
\begin{equation}
\dot{q} (y,t) =  I(t)^2 \frac{\rho_e }{S_e^2} , \label{loadheat}
\end{equation}
with $\rho_e$ the temperature-independent resistivity of the aluminum layer and $S_e$ its cross section. The boundary conditions are:
\begin{eqnarray*}
T(y=0,t) & = & T_0 ,   \\
\kappa[T(y=h,t)] S_e \, \frac{\partial T(y=h,t)}{\partial y} & = & \frac{\dot{Q}_0}{2} , 
\end{eqnarray*}
reflecting the thermalization to the bath $T_0$ at $y=0$, and the heat load $\dot{Q}_0= I(t)^2 \rho_e l / S_e $ due to the paddle at $y=h$.
Rewriting Eq. (\ref{heat}) after injecting Eq. (\ref{loadheat}) and the metal properties (parameters $\alpha, \gamma$), we obtain: 
\begin{equation}
\frac{1}{2} \rho_m \, \gamma \frac{\partial T(y,t)^2}{\partial t} -\frac{1}{2} \alpha \frac{\partial^2 T(y,t)^2}{\partial y^2}  = \frac{1}{2} I_{heat}^2 \frac{\rho_e }{S_e^2} \left( 1+ \cos \left[ 2 \omega'' t \right] \right)  \nonumber .  
\end{equation}
The above equation is analytically solvable if we chose :
\begin{equation}
 T(y,t)^2 = Re \left[ A_0 (y) + \Delta A(y) \exp (+i 2 \omega'' t)  \right] \nonumber
\end{equation}
written in complex form. The two functions $A_0 (y)$ and $\Delta A(y)$ are now solutions of:
\begin{eqnarray}
-\frac{1}{2} \alpha \frac{\partial^2 A_0 (y) }{\partial y^2} & = & \frac{1}{2} I_{heat}^2 \frac{\rho_e }{S_e^2} , \label{static} \\
-\frac{1}{2} \alpha \frac{\partial^2 \Delta A (y) }{\partial y^2} + i \, \rho_m \gamma \, \omega'' \Delta A(y) & = & \frac{1}{2} I_{heat}^2 \frac{\rho_e }{S_e^2} \label{tdep} .
\end{eqnarray}
with boundary conditions:
\begin{eqnarray*}
A_0      (y=0) & = & T_0^2 , \\
\Delta A (y=0) & = & 0 , \\
\frac{1}{2} \alpha \, S_e \frac{\partial A_0 (y=h) }{\partial y} & = &  \frac{1}{4} I_{heat}^2 \frac{\rho_e l}{S_e} , \\
\frac{1}{2} \alpha \, S_e \frac{\partial \Delta A (y=h) }{\partial y} & = &  \frac{1}{4} I_{heat}^2 \frac{\rho_e l}{S_e} .
\end{eqnarray*}
We finally obtain:
\begin{eqnarray*}
A_0      (y) & = & \frac{I_{heat}^2 \rho_e}{2 S_e^2 \alpha} \left[ -y^2+y\left(2h+l\right) \right] + T_0^2, \\
\Delta A (y) & = & \frac{I_{heat}^2 \rho_e}{2 S_e^2 \alpha} \, d^2 \left( 1 - \exp \left[-(1+i) \frac{y}{d} \right] \right) \times \\
& & \!\!\!\!\!\!\!\!\!\!\!\!\!\!\!\!\!\!\!\!\!\!\!\!\!\!\!\!\!\!\!\! \left(\frac{ +i \left[ \exp \left(+[1+i] \frac{y}{d} \right)-\exp \left(+[1+i] \frac{2h}{d} \right) \right]+\frac{1}{2} (1-i) \frac{l}{d} \left[ \exp \left(+[1+i] \frac{h}{d} \right) + \exp \left(+[1+i] \frac{h+y}{d} \right)\right] }{1+\exp \left(+[1+i] \frac{2h}{d} \right)}\right),
\end{eqnarray*}
where we introduced a new lengthscale $d=\sqrt{\alpha/(\gamma \rho_m \omega'')}$. Note that in the limit $\omega'' \rightarrow 0$ we recover $\Delta A (y) = \frac{I_{heat}^2 \rho_e}{2 S_e^2 \alpha} \left[ -y^2+y\left(2h+l\right) \right]$ as we should. For finite $\omega''$, the AC heating term $\left|\Delta A (y)\right|$ is always smaller than $A_0 (y)-T_0^2$, and finally becomes negligible for $\omega'' \rightarrow +\infty$.

In practice, the lock-in measurement is fairly slow (typically a second per point, and about 15$~$min. per sweep).
For a finite $\omega''> 2 \pi \, 10~$Rad/s, the obtained resonance essentially filters-out the thermal AC component $\Delta A$.
In the following thermo-mechanical modeling, we can thus limit ourselves to the static DC term $A_0$: when describing our experimental results, the heat capacity of the beam turns out to be irrelevant, and only $\alpha$ (i.e. conductivity) matters.

Knowing the temperature distribution along one foot of the structure, we built a model integrating it out and quantitatively predicting the observed average mechanical parameters. The potential energy and the dissipated energy per vibration cycle can be written as:
\begin{eqnarray*}
E_p & = & \frac{1}{2} \int^{h}_{0} \left[ \frac{\partial^2 f(y,t)}{\partial y^2} \right]^2 E[T(y,t)] I_y \, dy, \\
P_d & = & -2 \int^{h}_{0} \frac{d \Lambda }{dy}[T(y,t)] \left[ \frac{ \partial f(y,t)}{\partial t} \right]^2 \, dy,
\end{eqnarray*}
with $E(T)$ the effective Young modulus of the composite beam and $d\Lambda/dy(T)$ the dissipation constant per unit length. Both temperature variations originate in the metal contribution \cite{coatings}. The function $f(y,t)$ represents the mode shape, and $I_y$ the second moment of area of the beam. 
From Fig. \ref{temps} we realize that a fairly reasonable estimate of $E(T)$ and $d\Lambda/dy(T)$ can be produced by a linear fit (dashed lines). To compute the above integrals, we use the Rayleigh approximation for the mode shape and write $f(y,t) = x(t) \, f_R(y)$ with $f_R(y)=3/2 (y/h)^2-1/2 (y/h)^3$. The calculation brings:
\begin{figure}[t!]
\includegraphics[height=9 cm]{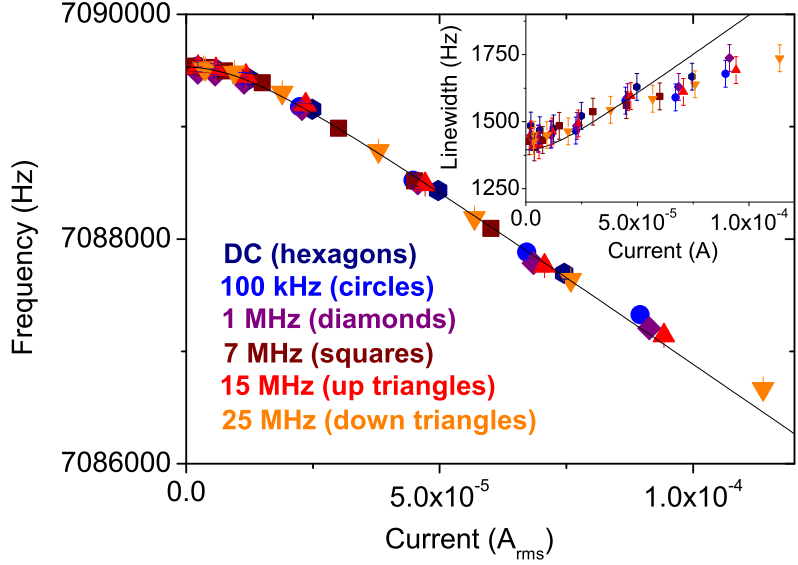}
\caption{\label{freqTforI} (Color online) Frequency of the resonance as a function of the heating current. The different symbols (colors) stand for different frequencies $\omega''$ (see Figure). Inset: same results for the linewidth. The black lines are theoretical predictions developed in Sec. \ref{theorthermal}. Note that the deviation above 70$~\mu$A is generated by the temperature of the structure exceeding 15$~$K. Error bars $\pm 150~$Hz on the frequency and $\pm50~$Hz on the linewidth.}
\end{figure}
\begin{eqnarray*}
k_0(T) & = & k_0(T_0) \left( 1+ \frac{dE/dT}{E(T_0)} \left[\frac{\int^{h}_{0} f_R''(y)^2 \frac{\Delta T_{DC} (y)}{\Delta T_{max}} dy}{\int^{h}_{0} f_R''(y)^2  dy}\right] \Delta T_{max} \right), \\
\Lambda(T) & = & \Lambda(T_0) \left( 1+ \frac{d \! \left(\frac{d\Lambda}{dy}\right)\!/dT}{\frac{d\Lambda}{dy}(T_0)} \left[\frac{\int^{h}_{0} f_R(y)^2 \frac{\Delta T_{DC} (y)}{\Delta T_{max}} dy}{\int^{h}_{0} f_R(y)^2  dy}\right] \Delta T_{max} \right),
\end{eqnarray*}
with the usual definitions $\omega_0(T)=\sqrt{k_0(T)/m_0}$ (resonance frequency) and $\Delta \omega (T) =2 \Lambda (T)/m_0$ (linewidth). We have $\Delta T_{DC} (y) = \sqrt{\frac{I_{heat}^2 \rho_e}{2 S_e^2 \alpha} \left[ -y^2+y\left(2h+l\right) \right]+T_0^2}-T_0$ and $\Delta T_{max}=\Delta T_{DC}(y=h)$ by definition. In our experiments $T_0=4.2~$K, and the slopes $dE/dT$ and $d \! \left(\frac{d\Lambda}{dy}\right)\!/dT$ are obtained from the caption, Fig. \ref{temps}. The terms in brackets are adimensional parameters that we estimate numerically. In our experiments, we obtain for the frequency a number between $0.296$ and $0.430$, and for the linewidth between $0.850$ and $0.910$ (from small to large powers, respectively).
Note that the modeling, which is valid up to about 15$~$K, has no free parameters.

\subsubsection{Magnetomotive line $G_I(\omega)$}
\label{magneto}

For the magnetomotive actuation, a current $I(t)$ is fed through the metallic layer via a cold bias resistor $R$ and an applied voltage $\tilde{V}_I(t)$. 
Since the generators have a finite output impedance $Z_I(\omega)$,$Z_V(\omega)$, the actual bias impedance $R_{equiv}$ that generates the current $I=\tilde{V}_I/R_{equiv}$ is (see Fig. \ref{setup}):
\begin{equation} 
R_{equiv} = \left( R + r\frac{1 +i \, Z_V(\omega) C_0 (\omega) \omega }{1 + i \, [ r+Z_V(\omega)] C_0 (\omega) \omega  } \right) \left(1 + \frac{i \, r  C_0 (\omega) \omega  }{1+i \, Z_V(\omega)C_0 (\omega) \omega   } \right) .
\end{equation}
In the above, the high input impedance of the detection $Z_D(\omega)$ has been taken to infinity (i.e. 1$~$M$\Omega$, 20$~$pF coupling).
In practice, the frequency-dependent corrections to $R+r$ are small, and are all incorporated to the line transmission coefficient $G_I(\omega)$. We thus chose for simplicity $R_{equiv} \equiv R+r$ in the whole of the paper.
\begin{figure}[t!]
\includegraphics[height=9 cm]{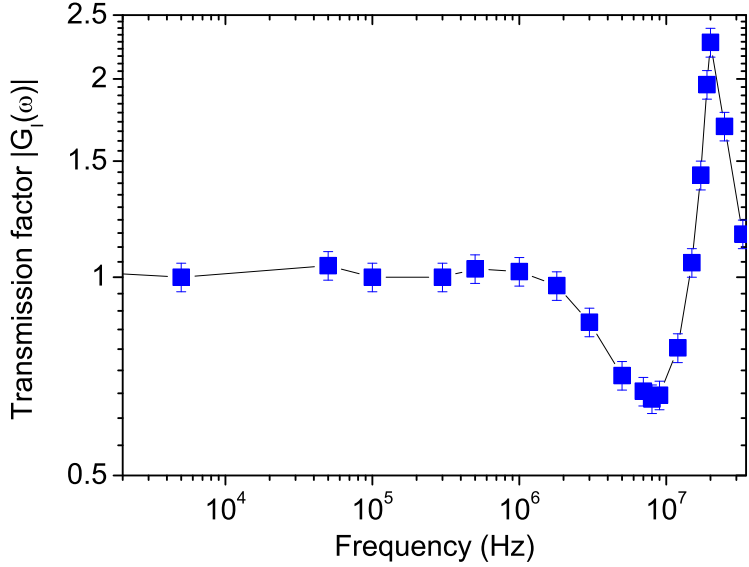}
\caption{\label{GI} (Color online) Transmission coefficient $\left|G_I(\omega)\right|$ as a function of the frequency $\omega/(2 \pi)$. Error bars about $\pm 5~$\%.}
\end{figure}

In Fig. \ref{freqTforI} we present the frequency of the resonance as a function of the injected heating current $I_{heat}$. To obtain this figure, we write (in complex form) $I_{heat}(t)=Re[\tilde{V}_{heat}/(R+r) \exp(+i\omega'' t)]$ and $\tilde{V}_{heat} = V_{heat} \, G_I(\omega'')$. For each heating frequency $\omega''$, we scale the $x$-axis (heating current) on the DC result. All the data collapse on a single curve, well described by the theoretical analysis presented on Sec. \ref{theorthermal} (black lines). The parameters used for the theoretical prediction are consistent with the experimentally defined properties of the NEMS within typically 10$~$\%.
Above about 70$~\mu$A, deviations start to be visible, since the temperature of the NEMS raises above 15$~$K: the heating becomes less efficient because the metallic layer begins to be thermally short-circuited by the heat conduction of the silicon.

The obtained scaling coefficients produce the transmission curve $\left|G_I(\omega)\right|$ presented in Fig. \ref{GI}. As expected, the low frequency region tends to 1, while above 1$~$MHz the imperfections of the line start to show: we clearly distinguish a first low-pass tendency (distributed $RC$) followed by a sharp line resonance. Although fitting theoretically this result is impossible because we lack an exact description of the wiring, a {\it PSpice}$\textsuperscript{\textregistered}$ \cite{spice} simulation qualitatively reproduces the data.

\subsubsection{Capacitive line $G_V(\omega), C_0(\omega)$}
\label{capalines}

When an AC voltage $\tilde{V}_G$ is applied to the gate electrode, a current $I(t)$ flows through the NEMS. Since the generators have a finite output impedance $Z_V(\omega)$,$Z_I(\omega)$, the actual coupling capacitance $I=C_{equiv}\,d\tilde{V}_G/dt$ writes, in complex form (see Fig. \ref{setup}):
\begin{equation} 
\frac{1}{i\, C_{equiv}\, \omega} = \frac{1}{i\,C_0 \,\omega} \left( 1+ \frac{i \,r C_0(\omega)  \omega}{1+\frac{r}{R+Z_I(\omega)}}\right) \left( 1+ \frac{r}{R+Z_I(\omega)} \right)  .
\end{equation}
Again, the lock-in input impedance $Z_D(\omega)$ has been assumed to be infinite. Similarly to the preceding paragraph, we realize that the corrections to $C_0$ are small, and we will incorporate them in the definition of $C_0$ itself, writing simply $C_{equiv} \equiv C_0$.

\begin{figure}[t!]
\includegraphics[height=9 cm]{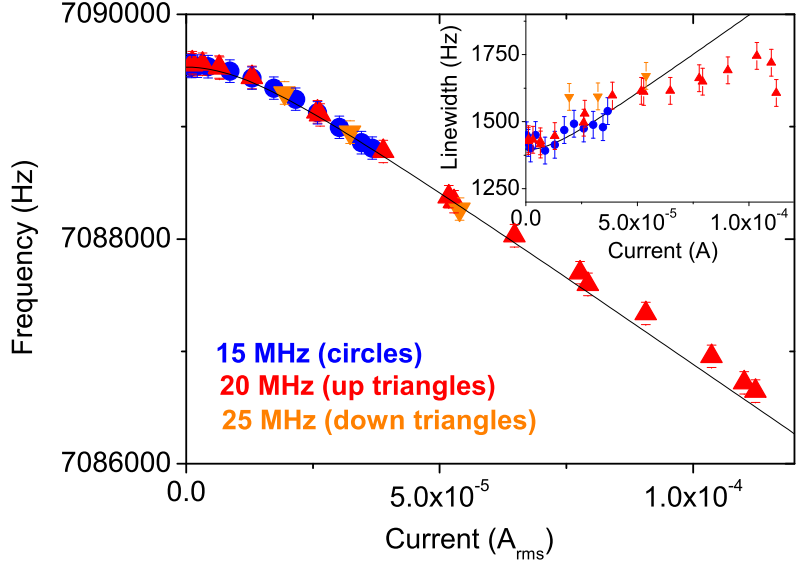}
\caption{\label{freqTforV} (Color online) Same graph as Fig. \ref{freqTforI} obtained with a heating current generated by a gate voltage drive. Only above a few MHz the heating effect is resolved. }
\end{figure}

In Fig. \ref{freqTforV} we present the frequency shift due to the heating effect. Since the frequency increases quadratically with the gate voltage $\tilde{V}_G$, Fig. \ref{freq},  we first subtract a quadratic dependence $[V_{heat} \, \left|G_V(\omega'')\right| ]^2$, defining thus the transmission factor $\left|G_V(\omega'')\right|$. The remaining downwards frequency shift is thus solely due to the current $C_{0}(\omega'')\,d\tilde{V}_{heat}/dt$. The higher the frequency, the larger the current is and thus the more efficient the heating effect becomes. By scaling the $x$-axis (heating current) of these curves on the heating data of Sec. \ref{magneto}, Fig. \ref{freqTforI}, we finally extract $C_{0}(\omega'')$.
\begin{figure}[t!]
\includegraphics[height=8.55 cm]{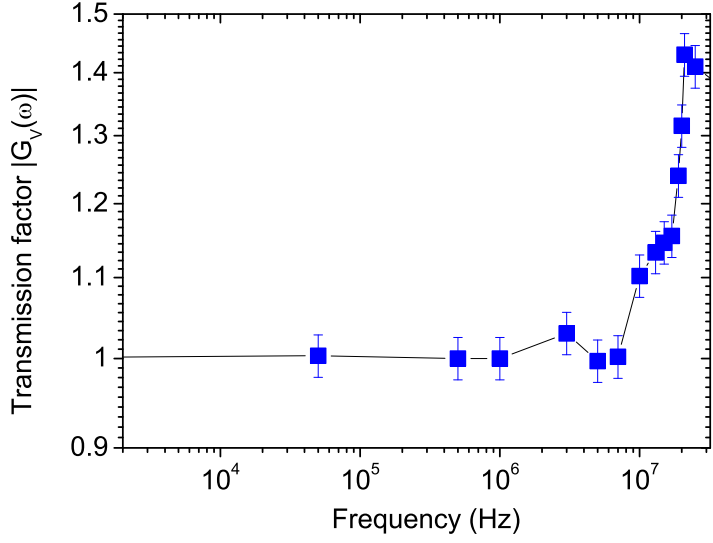}
\caption{\label{GV} (Color online) Transmission coefficient $\left|G_V(\omega)\right|$ as a function of the frequency $\omega/(2 \pi)$. Error bars typically $\pm 2.5~$\%.}
\end{figure}
\begin{figure}[t!]
\includegraphics[height=8.55 cm]{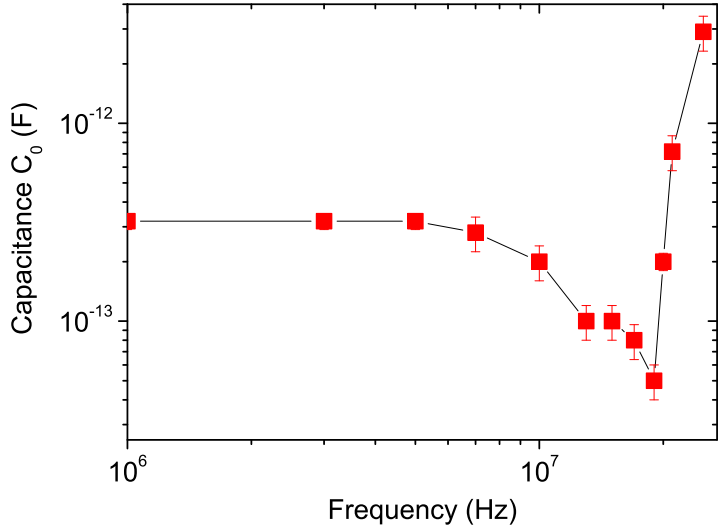}
\caption{\label{capaFreq} (Color online) Coupling capacitance $C_0(\omega)$ as a function of the frequency $\omega/(2 \pi)$. The low frequency data correspond to the measured value $C_{00} = 0.32~$pF. The largest error bars are $\pm 20~$\%.}
\end{figure}

The obtained transmission curve $\left|G_V(\omega)\right|$ is shown in Fig. \ref{GV}. Again, the low frequency region tends to 1 while a clear line resonance is seen around 20$~$MHz. 
In Fig. \ref{capaFreq} we give the coupling capacitance $C_0(\omega)$. Only above about a few MHz the heating effect starts to be visible. The low frequency value thus corresponds to the one measured with the lock-in amplifier (Sec. \ref{explain}, $C_{00} = 0.32~$pF). At higher frequencies, the thermally measured parameter roughly follows the same tendencies as the transmission coefficients $\left|G_I(\omega)\right|$, $\left|G_V(\omega)\right|$ (Figs. \ref{GI} and \ref{GV}). The error bars are quite large, due to the $\tilde{V_G}^2$ quadratic contribution subtraction procedure.
A {\it PSpice}$\textsuperscript{\textregistered}$ \cite{spice} simulation can be used to qualitatively reproduce these results.

\subsubsection{Detection line $G_D(\omega)$}
\begin{figure}[t!]
\includegraphics[height=9 cm]{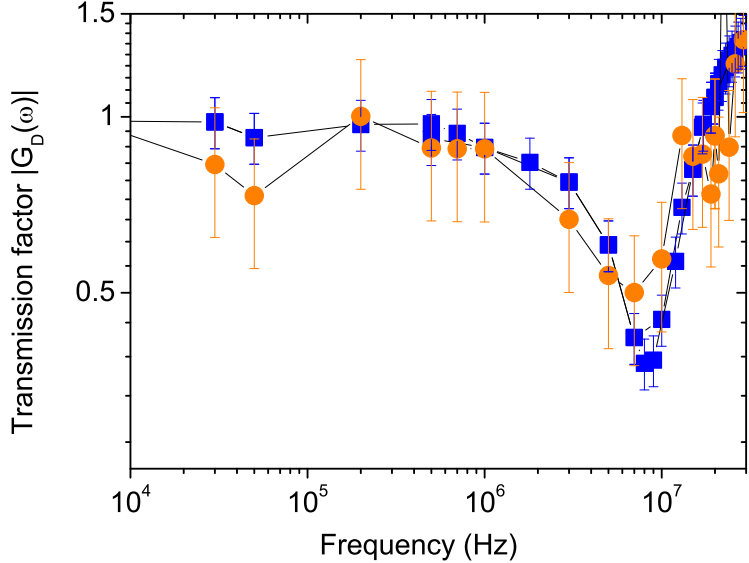}
\caption{\label{GD} (Color online) Transmission coefficient $\left|G_D(\omega)\right|$ as a function of the frequency $\omega/(2 \pi)$. The two symbols correspond to measurements performed via the magnetomotive port (squares), and via the gate capacitive port (circles). Error bars $\pm 10~$\% and $\pm 25~$\% respectively.}
\end{figure}

Knowing the input ports transmission coefficients $\left|G_I(\omega)\right|$, $\left|G_V(\omega)\right|$, we {\it directly} measure the transmission $\left|G_D(\omega)\right|$ of the detection line without resorting to a mechanical property of the device. 
We inject a small current through the NEMS metallic layer $I(t)$ at frequency $\omega$, keeping the amplitude small enough to avoid heating effects. 
This can be done either by the magnetomotive port (on the bias resistor $R$), or by the gate port (using the capacitive coupling $C_0$).
The ohmic voltage $r I(t)$ appearing at the NEMS ends is then measured as a function of the frequency $\omega$.

The resulting curves are shown in Fig. \ref{GD} for both current-feeding procedures. The capacitively obtained curve has clearly less resolution because of our limitation on the measurement of $C_0(\omega)$. The same features as on $\left|G_I(\omega)\right|$, $\left|G_V(\omega)\right|$ and $C_0(\omega)$ can be seen, 
in qualitative agreement with a {\it PSpice}$\textsuperscript{\textregistered}$ \cite{spice} simulation.

\section{CONCLUSIONS}

In the present paper we report on experiments performed on a nanomechanical oscillator (NEMS) in cryogenic vacuum at Helium temperatures. The device has two actuation ports, one being a (linear) magnetomotive drive and the other one a (nonlinear) capacitive drive. The detection of the motion is performed via the measurement of the voltage induced by the magnetic flux modulation. We illustrate on our device new techniques enabling the full {\it in situ} characterization of the device, based only on electric measurements. \\
In a first part, we demonstrate that we are able to reconstruct experimentally the coupling capacitance $C(x)$ from its Taylor series expansion. Each coefficient is obtained by fitting a corresponding NEMS property that is controlled by the gate voltage bias $V_G$.\\ 
In a second distinct part, we use  thermal properties of the NEMS (which are essentially DC) to carefully calibrate the bias lines down to the moving part, defining thus displacements and forces in absolute S.I. units (meters and Newtons respectively). We present a thermal model validating our results. The model has no free parameters, but the method itself {\it does not} require its knowledge. \\
Finally, from the characteristic displacement versus force measurement we can deduce separately the spring constant and the mass of the mode under study. \\
We believe that the technique can be extremely useful for a variety of experiments using nanomechanical resonators, since the calibration procedure does not require any particular qualities for the connecting lines and can be straightforwardly adapted to a variety of devices. In particular, high impedance environments are perfectly tolerated, even in the $10~$MHz range.

{\it Note added:} We used this technique to calibrate the device used in experiments described in Refs. \cite{PRBtobe,APLtobe}. We are currently using it in experiments mounted in a $^4$He cryostat and {\it also} in a dilution unit. The calibration procedure is then performed {\it before} starting the unit.

\begin{acknowledgments}

We wish to acknowledge the support of Thierry Fournier, Christophe Lemonias, and Bruno Fernandez in the fabrication of samples. 
We greatly thank Olivier Exshaw and Julien Minet for help with the electronics equipment, 
together with Christophe Guttin for help in {\it PSpice} $\textsuperscript{\textregistered}$ \cite{spice} simulations of the setup.
The authors wish to acknowledge valuable discussions with Thierry Fournier, Jeevak Parpia and Vincent Bouchiat. 
We acknowledge the support from MICROKELVIN, the EU FRP7 low temperature infrastructure grant 228464 and of the 2010 ANR French grant QNM n$^\circ$ 0404 01.

\end{acknowledgments}

%
%

\end{document}